\documentclass[submission]{eptcs}
\providecommand{\event}{E. Formenti (Ed.): AUTOMATA \& JAC 2012} 
\usepackage{subfigure}

\usepackage{epsfig,graphicx,subfigure,latexsym}

\newtheorem{theorem}{Theorem}

\newtheorem{lemma}{Lemma}

\newtheorem{property}{Property}

 \newcommand{\pf}{{\sc Proof} }
\newcommand{\ep}{~{\vrule height4pt width4pt depth2pt}\\}

\newcounter{smallitemizec}

 \title{ On the Parity Problem in  One-Dimensional\\
  Cellular Automata}
  \author{Heather Betel
  \institute{School of Electrical Engineering and Computer Science\\  University of Ottawa\\ Ottawa, Canada }
  \email{hbetel@site.uottawa.ca}
  \and  Pedro P.B.~de Oliveira
  \institute{Universidade Presbiteriana Mackenzie\\ Faculdade de Computa\c{c}{\~a}o e Inform{\'a}tica\\ S\~{a}o Paulo,  Brazil}
  \email{pedrob@mackenzie.br}
  \and  Paola Flocchini
  \institute{School of Electrical Engineering and Computer Science\\  University of Ottawa\\ Ottawa, Canada }
  \email{flocchin@site.uottawa.ca}
}
\def\titlerunning{One-Dimensional Parity Problem}
\def\authorrunning{H. Betel, P. de Oliveira \& P. Flocchini}
\begin{document}
 
\maketitle 
 
\begin{abstract}
We consider the parity problem in one-dimensional, binary, circular cellular automata:
if the   initial configuration contains an odd number of 1s, the lattice  should converge to all 1s; otherwise,
 it should converge to all 0s.  It is easy to see that the problem is ill-defined for even-sized lattices 
 (which, by definition, would never be able to  converge  to 1). 
 We then consider only odd lattices.
 
We are interested in determining  the minimal neighbourhood that allows
the problem to be solvable  for any  initial configuration. 
On the one hand, we show that radius 2 is not sufficient, 
proving that there exists no radius 2 rule that can possibly 
solve the parity problem from arbitrary initial configurations.  
On the other hand, we design a radius 4 rule that converges 
correctly for any initial configuration and we formally prove its correctness.
Whether  or not there exists a   radius 3 rule that solves the parity problem 
 remains an open problem.
\end{abstract}

\section{Introduction}

Understanding the nature of computations within cellular automata 
remains an elusive problem. In fact, in spite of their long-proclaimed ability 
to perform computations, very  little is still known 
as to how we should design the local state transitions towards achieving 
a given global behaviour.
As examples are designed or found by search, it is inevitable to try to understand their 
underlying programming  language; but the truth is, to this date, every attempt along these lines has 
fallen into the strenuous effort of 
trying to tame local state patterns towards the global state target, or trying to make sense of the latter 
in terms of the former \cite{Griffeath.Moore.2003}.

On the other hand, studying how to   employ  local actions  
to achieve  desirable global behaviours is of utmost importance and extensively investigated 
in many other evolving systems    (e.g., distributed systems, 
mobile robots, population protocols).
In such systems, in fact, understanding  the limitations and the power of local interactions to solve
global computations  has immediate   implications on the design of efficient and scalable 
  solutions (e.g., see \cite{Angluin.1980,Angluin.2007,Lenzen.2008,Peleg.2000}).  
Cellular automata (CAs) are the simplest possible evolving systems, and understanding 
the impact that neighbourhood size has on computability    could have consequences  
for more complex systems based on local interactions.

This paper is aligned with these efforts.  Here, we concentrate on the one-dimensional {\em parity problem}, 
which has essentially the objective of figuring out the parity of an arbitrary binary string, by means of a one-dimensional, binary cellular automaton \cite{Sipper.1998}. The parity problem is a well-known benchmark task in various areas of computer science, typically camouflaged under the XOR operation on a binary input, as in artificial neural networks \cite{Haykin.2008}, but it also lends itself to the context of cellular automata, as a typical case of a global problem that has to be solved by purely local processing.
The problem is formulated under periodic boundary conditions and arbitrary finite 
lattice size, so that, if the parity of the global configuration is odd, the lattice is supposed to lead to an homogeneous configuration with only 1s; otherwise, it should converge to all 0s \cite{Lee.Xu.Chau.2001}.

The notion of parity has appeared quite often in the CA literature, even if implicitly, as it bears relevance to the related notion of additivity of CA rules \cite{CCNC.1997,Voorhees.2009}. However, the parity problem per se has not been extensively studied, particularly in comparison  with the well-known benchmark CA task of {\em density classification}, where the aim is to determine the most frequent bit in the initial configuration of an odd-sized lattice,
 also by reaching an homogeneous configuration. The density classification problem, in fact, 
has been extensively investigated and fully understood  in odd-sized lattices. 
In particular, it has been shown that there exists no single rule able to solve the problem for any arbitrary 
initial configuration. Combinations of rules have been devised, however, as well as probabilistic solutions to the problem
 (e.g., see  \cite{deoliveira.2006,Fates.2011,Fuks.1997,Wolz.deOliveira.2008}).

An advantage in favour of the parity problem is that, from the perspective of automata theory, 
it is simpler than its kin, insofar as the notion of parity can be handled by finite automata, 
whereas the ability to compare arbitrarily variable quantities (which is inherent to density 
classification) requires at least a pushdown automaton \cite{Hopcroft.Motwani.Ullman.2006}. 
In fact, the increased simplicity of the parity problem is reflected in the fact that it is easier to find good rules for it, 
by searching, than for density classification \cite{Wolz.deOliveira.2008}. Therefore, there are strong reasons 
for considering the parity problem generally more tractable and amenable to analysis, which makes it 
a serious candidate for case studies that might help the understanding of the nature of computation in CAs in general.

The parity problem is ill-defined for even-sized lattices (by definition, an all 1 configuration
converges to an all 0 configuration making it impossible for any rule to converge to 1). 
 Modifying the definition of the problem to allow the target homogeneous configuration  to be achieved only once, 
 and not as a fixed point,  the problem   becomes solvable also for even lattices. In fact, by relying on this variation, 
 it can be perfectly solved by a carefully engineered sequence of rule applications, quite surprisingly, of elementary cellular automata \cite{Martins.deOliveira.2009}.  However, if we do not want to change the definition of the problem, 
 it is then necessary  to restrict the  study to odd-sized lattices. 
  We then say that a  CA  rule is     {\em perfect } if   it  solves the parity
  problem for     arbitrary  initial odd-sized configurations.  

 Unlike the  density classification problem, we show that the parity
  problem can, indeed,  be solved by a single rule. Besides being interested in 
 its general solvability, we are also interested in determining the 
 minimal neighbourhood that allows the construction of a perfect rule.
With this goal in mind, we first prove   that radius 2 is not sufficient for a perfect rule to exist. 
We first identify several constraints to which such a perfect rule is subject and
we show that no rule is feasible with all of these constraints.
  We then show that the problem becomes  solvable when CAs have 
  radius 4: our proof is constructive 
as we design a perfect rule and we prove its correctness.
We leave open the case of radius 3, for which there is strong  empirical evidence that no 
perfect rule exists, but that there might be radius 3 rules that would solve the problem for 
prime-sized lattices \cite{Wolz.deOliveira.2008}.

\section{Notation and Basic Facts}

  We consider one-dimensional, binary  cellular automata (CAs) on finite lattices 
 with periodic boundary conditions. 
  Let  $f:\{0,1\}^{2r+1} \rightarrow \{0,1\}$ denote the local rule of a  CA with radius $r$.
 The global dynamics of a one-dimensional cellular automaton composed of $n$ cells and radius $r$
  is then defined by the global rule (or transition function):
 $F:\{0,1\}^{n} \rightarrow \{0,1\}^{n}  \,\,\,\,\,\,\, s.t. \,\,\,\,\,\, 
 \forall {\bf X}  \in \{0,1\}^{n},
 \forall i \in \{0,\ldots, n-1\},F(X)_i = f(x_{i-r} \ldots,x_i,\ldots, x_{i+r})$, where all operations
 on indices are modulo $n$.
 
A  {\em fixed point}  ${\bf P} \in \{0,1\}^n$  of a circular
CA with global transition  rule $F$
is a configuration ${\bf P}$ such that $F({\bf P}) = {\bf P}$.

  We say that a cellular automaton {\em converges to a configuration $P$} from configuration $X^0$
   if  $P$ is a fixed point and if for some finite $n$, $F^n(X^0)=P$ where $F^n$ is the $n^{\textrm {th}}$ iteration of $F$. We are particularly interested in the homogeneous configurations as fixed points and will refer to these as the {\em 0-configuration} and the {\em 1-configuration}.

We say that a local rule solves the {\em parity problem}  if,  
 starting from an arbitrary initial configuration, on an arbitrarily sized lattice, 
 the cellular automaton converges to the 0-configuration, if and only if the initial configuration contains 
 an even number of 1s,  and converges to the 1-configuration otherwise.

 Since a rule solving the parity problem must converge to the homogeneous configurations, we have our first two properties, of perfect rules.
 
 \begin{property} If $f$ solves the parity problem,
  then $f(0,\cdots,0)=0$ and $f(1,\cdots,1)=1$.\label{fixed}\end{property}

\noindent   It is immediately obvious that,
 by definition, no solution exists for even-sized lattices. 
  \begin{theorem} 
Consider circular CAs with radius $r$ and even size $n$. 
There exists no rule that   works correctly from any initial configuration.
 \end{theorem}
\pf
Trivially $f(1\ldots 1)=0$ otherwise the configuration with all 1s would incorrectly
converge to 1.
Since $(11111\ldots 111)$ is not a fixed point, it follows that no initial configuration can ever converge to the 1-configuration.
\ep

\noindent For this reason, from now on,  we consider only {\em odd}-sized lattices and we call  
a  rule {\em perfect} if it solves the parity problem for any   odd-sized lattice, starting from any initial configuration.

\begin{figure}[!h]
 \centering     
 \mbox{  {\scalebox{.60}
{\includegraphics{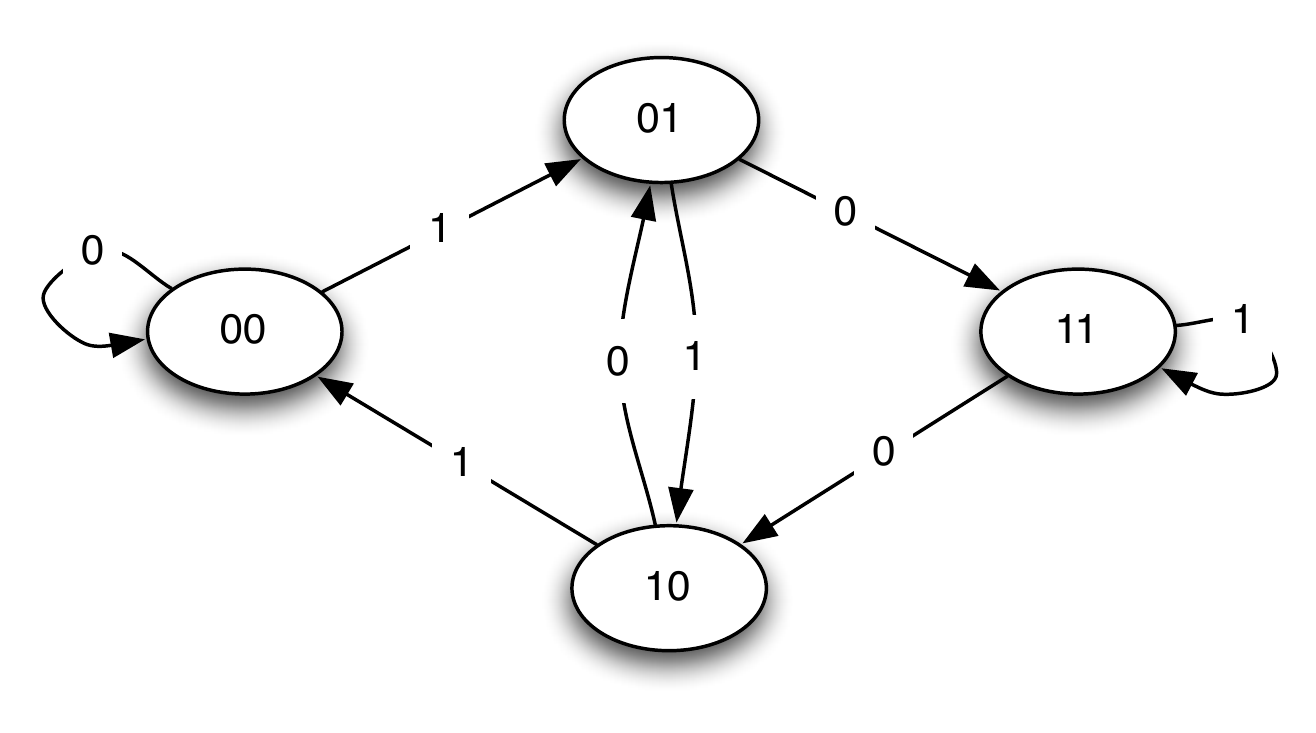}}}
}
\caption{De Bruijn graph for the local parity rule (150).}\label{rad2}
\end{figure}

We now recall the definition of  de Bruijn graphs, which  
are useful tools for representing CA rules and which will be used
in the subsequent   sections.
The de Bruijn graph of a local rule of radius $r$  is a directed graph on $2^{2r}$ nodes, 
one for each value in the set $\{0,1\}^{2r}$.  
There is an edge from node $x_0\dots x_{2r-1}$ to node $y_1\dots y_{2r}$ if $x_i=y_i$ for all $i$ from 1 to $2r-1$.  
These edges are labelled with the value of the local function at $(x_0,\dots, x_{2r-1},y_{2r})$, $f(x_0,\dots, x_{2r-1},y_{2r})=f(x_0,y_1,\dots, y_{2r})$.  
Note that the shape of the de Bruijn graph for a local rule of a given family (i.e., those with the same neighbourhood and states) is fixed, only the edge labels change. 
For example, Figure \ref{rad2} shows the de  Bruijn graph for a radius 1 rule.
 
 The following is a necessary condition for parity preservation.
 \begin{property} A rule   preserves the parity of a configuration if active transitions always come in pairs.  That is, given a local rule $f$ of radius $r$, and any configuration $(x_0,\dots,x_{n-1})$, the number of times that $f(x_{i-r},\dots,x_i,\dots,x_{i+r})\neq x_i$ is even. \end{property}\label{pairs}

  It is also very simple to see that no solution exists for elementary circular CAs (i.e., with radius $1$).
  
  \begin{theorem} 
There exists no perfect rule  for elementary CAs.
 \end{theorem}
\pf
From Property \ref{fixed}, for any perfect  rule, we must have $f(000)=0$ and $f(111)=1$. Now consider a configuration containing a single 1.  In order to both maintain parity and move towards convergence, we must have $f(100)=f(010)=f(001)=1$.  Similarly, from the singleton 0, we must have $f(110)=f(101)=f(011)=0$. So the only possible perfect rule is the local parity check (rule 150). However, it is easy to see that such a rule does not converge from several initial configurations, for example from $(\ldots 0001000 \ldots)$. 
  \ep

\section{Impossibility with Radius 2}

In this section we  show that    with radius 2 it is impossible to construct a perfect parity rule.  

Our aim is  to show, first,  several 
necessary transitions  for a perfect rule, and second,  
the existence of a limited set of   feasible pre-images for
 the two  final homogeneous  configurations.
Each possible pair of feasible pre-images  further  induces  necessary   transitions 
for   a perfect rule,  significantly reducing the space of possible perfect rules. 
We conclude the proof by running exhaustive tests to verify that
in this set  there  exists no perfect rule having the necessary transitions. 
We begin with a series of  lemmas  that force certain transitions to 0 or 1.

Consider the  de Bruijn graph  for radius 2 rules.
A pre-image of the final 0-configuration corresponds to  a cycle  
of odd size and even parity.
A pre-image of the final 1-configuration is a cycle  
of odd size and odd parity.
Let ${\cal B}_0 $ be the subgraph  containing only 
the edges corresponding to  transitions to 0 and ${\cal B}_1$
the subgraph containing the edges corresponding to  transitions to 1.

\begin{lemma}\label{1grow}
In a perfect rule, three or five of the following must transition to 1: \\
 $(10000), (01000), (00100), (00010), (00001)$. 
 \end{lemma}
\pf
 A configuration consisting of a single 1 must eventually converge to all 1s, hence the number of 1s in the configuration must increase.  Furthermore, in order to maintain parity, it must increase to an odd number.  The five configurations above are the only ones occurring at the local level that are not all 0, and therefore 3 or 5 of them must go to 1.\ep

\noindent  Similarly, 
 \begin {lemma}\label{0grow}
 In a  perfect parity rule, three or five of the following must transition to 0:\\
  $(01111), (10111), (11011), (11101), (11110)$.
 \end{lemma}

 \begin{lemma}\label{pre}Neither  ${\cal B}_0$  nor  ${\cal B}_1$   can  contain a cycle of even length and odd parity.
 \end{lemma}
 \pf A cycle of even length in either ${\cal B}_0$  or  ${\cal B}_1$ will become a sequence having even parity at the next iteration since it will be either all 0s or an even number of 1s, so this cycle itself will have changed parity. Assume that the de Bruijn graph of a rule $F$ admits such a cycle, let $C$ be such a cycle and let $P$ be a cycle of odd length passing through a node of $C$.  For the rule to be perfect, $F(P)$ must have the same parity as $P$. Now consider a new cycle $P'$ formed from $P$ by adding the cycle $C$ where $P$ passes through it.  Since $C$ has even length, $P'$ has odd length. Since $C$ has odd parity, $P$ and $P'$ have different parity.  However, since $F(C)$ has even parity, $F(P')=F(P)$, hence the parity of $P'$ has changed and $F$ cannot be perfect.\ep
 
 \begin{lemma} \label{10101}
In a perfect parity rule either:
$i)$ $ f(10101)=1$ and $ f(01010)=0$, or 
$ii)$  $ f(10101)=0$ and $ f(01010)=1$.
\end{lemma} 
\pf This is a direct consequence of Lemma \ref{pre} since   $ f(10101)=  f(01010)$ would imply the
existence of the  even cycle with odd parity $(1010,0101,1010,0101,1010,0101)$ either in 
${\cal B}_0$ or in ${\cal B}_1$.
\ep

\begin{lemma}\label{4zeros}
In a perfect parity rule, it is impossible to have   four or more consecutive 0s 
in a pre-image of the 0-configuration.
\end{lemma}
\pf  Let the pattern $(0000)$ be present in a pre-image $P$ of the 0-configuration. Then  the following neighbourhood configurations must also be present and must be transitioning to 0: $(10000)$ and $(00001)$.  By Lemma \ref{1grow}, we must then have the following configurations transitioning to 1, so they may not occur in $P$: $(01000)$, $(00100)$, $(00010)$.  Hence our group of four 0s must be both preceded and followed by at least two 1s, thus entailing that we have the necessary transition set $S= \{f(11000),f(00011),f(10000),f(00001)\} \rightarrow 0$. Consider now an initial configuration containing a single 1 surrounded by 0s.
From $S$, we have that    0001000 can only grow to 0011100,  
but, again from $S$, we have that, from 0011100, no growth is possible anymore, which is a contradiction.\ep

\noindent Analogously, we have:

 \begin{lemma}\label{4ones}
In a perfect parity rule, it is impossible to have four or more consecutive 1s
in a pre-image of the 1-configuration.
\end{lemma}

From Lemma \ref{pre}, any feasible pre-image of the 0-configuration (resp. 1-configuration) 
is either a simple odd cycle $c$ with even (resp. odd) parity,  or the composition 
of cycles not containing any even cycle of odd parity.

So,  to identify feasible pre-images for final configurations
 for lattice size $n$ in the de Bruijn graph, 
we have to  find at least one cycle 
of size $n$  to be labeled 0 and one to be labeled 1, having the property that they do not include:\\
$(i)$ the self-loops (which are forbidden by Lemmas \ref{4zeros} and \ref{4ones}\\
$(ii)$ the 2-cycle (0101,1010) (which is forbidden by Lemma \ref{10101}); and\\
$(iii)$ an even cycle with odd parity  (Lemma  \ref{pre}). 
  \smallbreak
 
 \noindent By inspecting all cycles of size 5, we obtain that:
 
 \begin{lemma}\label{5}
In a  perfect rule  at least one of these three cycles in the de Bruijn graph must transition to 1:\\
 $ B_1^5  = (0011,0111,1110,1100,1001 )$ {\em (corresponding to configuration: 00111)}\\
  $ B_2^5 =   (0000,0001,0010,0100,1000)$  {\em   (corresponding to configuration: 00001)}\\
  $B_3^5=  (0101,1011,0110,1101,1010)$  {\em   (corresponding to configuration: 01011)}\\
  and one of these must transition to 0: \\
  $ W_1^5 =$ $(0001,0011,0110,1100,1000)$  {\em  (corresponding to configuration: 00011)}\\
$ W_2^5 =$ $( 0111,1111,1110,1101,1011)$  {\em  (corresponding to configuration: 01111)}\\
$ W_3^5 =$ $(0010,0101,1010,0100,1001)$  {\em  (corresponding to configuration: 00101)}
 \end{lemma}
\pf
$ B_1^5$, $ B_2^5$  and $ B_3^5$  (resp.  $ W_1^5$ ,  $ W_2^5 $ and $ W_3^5 $) are the only cycles corresponding 
to  feasible pre-images for the 1-configuration
  (resp. 0-configuration)  for lattices of size 5,  
  which do not violate Lemmas  \ref{10101},  \ref{4zeros}, and \ref{4ones}.
\ep

Consider, now, lattices of size 7. All cycles of length 7 have been enumerated  
and the only cycles that do not contradict  Lemmas  \ref{pre}, \ref{10101}, \ref{4zeros},  and   \ref{4ones}
and correspond to   feasible pre-images of the 1-configuration are:

\begin{description}
\item $B_1^7=$ $(0000, 0001, 0011, 0111, 1110, 1100,1000)$ (configuration: 0000111)
\item $B_2^7=$ $(0001, 0011, 0110, 1101, 1010, 0100,1000)$ (configuration: 0001101)
\item $B_3^7=$ $(0001, 0010, 0101, 1011, 0110, 1100,1000)$ (configuration: 0001011)
\item $B_4^7=$ $(1001, 0011, 0110, 1100, 1001, 0010, 0100)$ (configuration: 1001100)
\end{description}

\noindent  Analogously, the only cycles which do not violate  Lemmas  \ref{pre}, \ref{10101}, \ref{4zeros},  and   \ref{4ones} and correspond to   feasible pre-images of the 0-configuration are:

\begin{description}
\item $W_1^7=$ $(0001, 0011, 0111, 1111, 1110, 1100, 1000)$  (configuration: 0001111)
\item $W_2^7=$ $(0010, 0101, 1011, 0111, 1110, 1100,1001)$  (configuration: 0010111)
\item $W_3^7=$ $(0011, 0111, 1110, 1101, 1010, 0100,1001)$  (configuration: 0011101)
\item $W_4^7=$ $(0110, 1100, 1001, 0011, 0110, 1101, 1011)$  (configuration: 0110011)
\end{description}

From simple observation, we can rule out some of these cycles and combinations of cycles.
\begin{lemma} A perfect rule of radius 2 cannot have $ W_1^5$ as a pre-image of the 0-configuration.\end{lemma}
\pf  Cycle $ W_1^5$  shares at least one transition in common with each of the possible pre-images of the 1-configuration of size 7.
For example, $ W_1^5$, $B_1^7$ and $B_2^7$ all share the edge $(0001, 0011)$ in the de Bruijn graph. 
 Cycles $ W_1^5$ and $B_3^7$ share $(0010,0101)$, and $ W_1^5$  shares $(0011,0110)$ 
 with $B_4^7$.\ep 

\begin{lemma} A perfect rule of radius 2 cannot have $ B_2^7$ as a pre-image of the 1-configuration.\end{lemma}
\pf  First, if $ B_2^7$ is a pre-image of the 1-configuration, then $ W_2^5$ is a pre-image of the 0-configuration of size 5 since $ B_2^7$ and $ W_3^5$ share $(1010, 0100)$.  Cycle $ B_2^7$ also has transitions in common with $ W_1^7$, $ W_3^7$, $ W_4^7$.  It has no common transitions with  $ W_2^7$, however,  $ W_2^7$ and  $ W_2^5$ together form a cycle, namely, $(0111,1110,1101,1011)$, in violation of Lemma \ref{pre}. \ep

\noindent Similarly, we can show,
\begin{lemma} A perfect rule of radius 2 cannot have $ B_3^7$ as a pre-image of the 1-configuration.\end{lemma}

\noindent In fact, we can now restrict to very well defined possible cases.
\begin{lemma}\label{moreconstraints}
 A perfect rule of radius 2 must have $ W_2^5$ as a pre-image of the  0-configuration, $ B_2^5$ as a\\ pre-image of the 1-configuration and either
\begin{itemize}
\item  $ B_1^7$ as a pre-image of the 1-configuration and $ W_4^7$ as a pre-image of the 0-configuration, or 
\item $ B_4^7$ as a pre-image of the 1-configuration and $ W_1^7$ as a pre-image of the 0-configuration.\end{itemize}\end{lemma}
\pf From the lemmas above, we know that the only possible pre-images of the 1-configuration of size 7 are $ B_1^7$ and $ B_4^7$. Cycle $ B_1^7$ has transitions in common with all possible pre-images of the 0-configuration except $ W_4^7$.  Of the 5-cycle pre-images of the 1-configuration, only $ B_2^5$ is compatible with $ W_4^7$. Of the possible pre-images of the 0-configuration of size 5, neither $ W_2^5$ nor $ W_3^5$ poses any conflict, however, $ W_3^5$ makes it impossible to have any cycles of length 3 going to the 1-configuration, so that all configurations having a period of size 3, (i.e. configurations of the form $(001001001\cdots 001)$ will fail to converge). The proof is analogous beginning with cycle  $ B_4^7$.\ep

\begin{figure}[tbh]
 \centering     
 \mbox{  {\scalebox{.45}
{\includegraphics{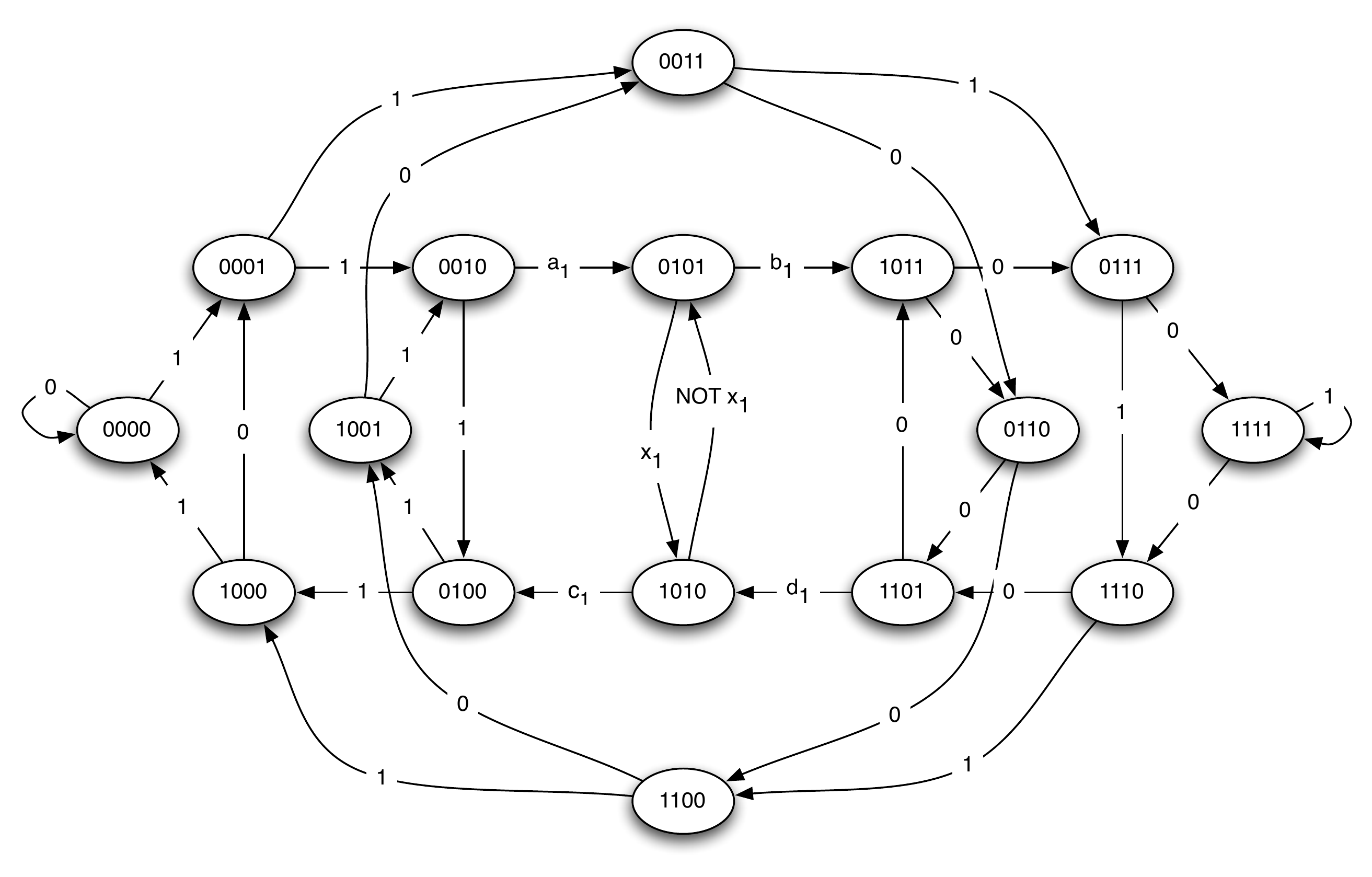}}}
}
\caption{Possible perfect rule for radius 2, with $ B_1^7$ and $ W_4^7$.}\label{r2v1}
\end{figure}

\begin{figure}[!h]
 \centering     
 \mbox{  {\scalebox{.45}
{\includegraphics{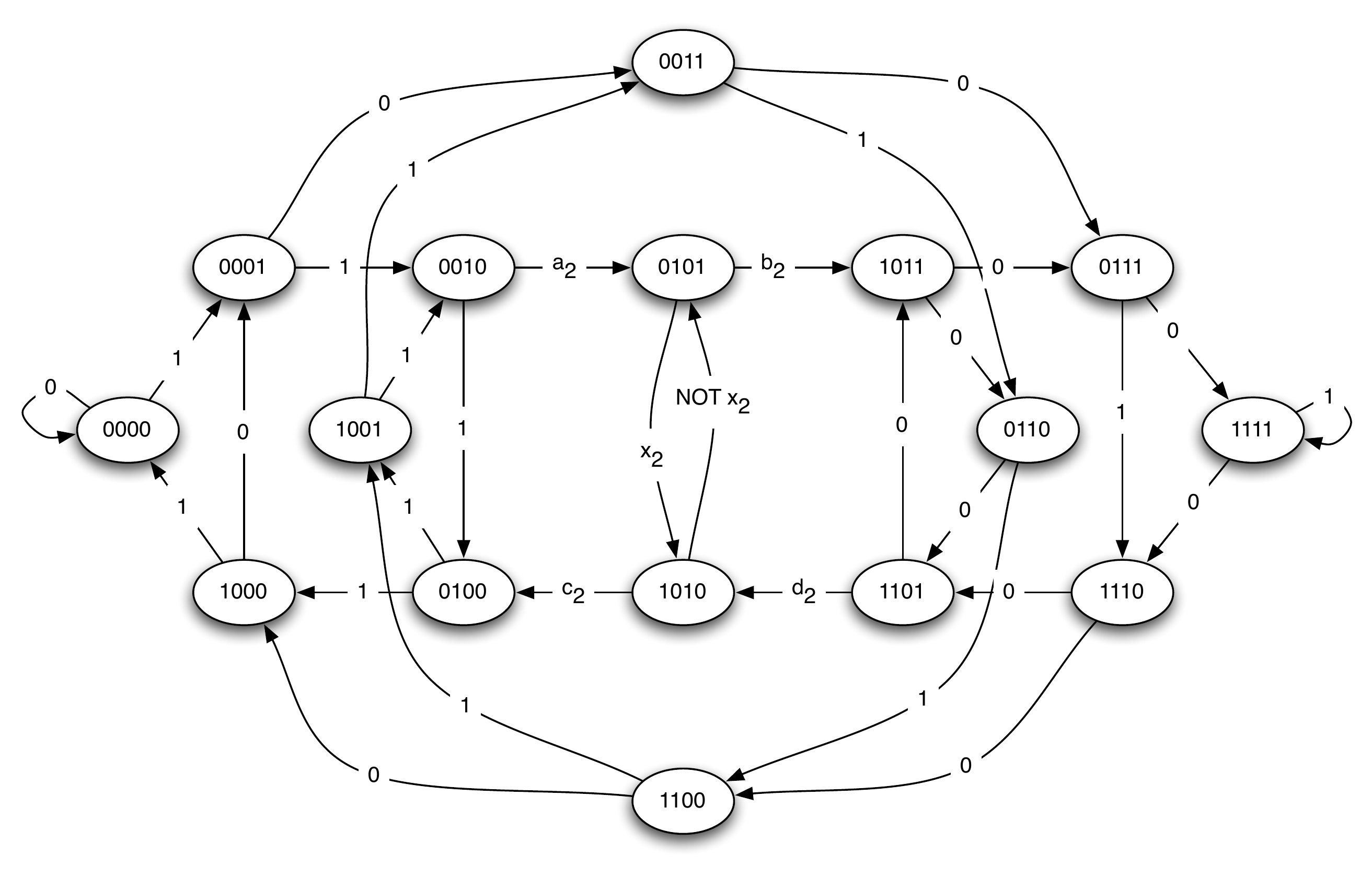}}}
}
\caption{Possible perfect rule for radius 2, with $ B_4^7$ and $ W_1^7$.}\label{r2v2}
\end{figure}

   \noindent We can finally conclude:
   
\begin{theorem} There is no perfect parity rule of radius 2.\label{rad2p}
\end{theorem}
\pf 
From the lemmas in this section it follows that for a perfect parity rule of radius 2, we must have one of the  de Bruijn graphs shown in Figure \ref{r2v1} or Figure \ref{r2v2}. 
Note that, in these graphs,  depending on the choice of $x_i$, we can place further restrictions on the remaining unknown edges 
 (again, due to the lemmas in this section).  Consider Figure \ref{r2v1}, if $x_1=1$, then in order to avoid creating cycles with an odd number of transitions, we must have $a_1\neq c_1, \ b_1\neq d_1$ from cycles $(0001, 0010, 0101,1010,0100,1000)$ and $(0101,1011,0111,1110,1101,1010)$. When $x_1=0$ by contrast, we must have $a_1= c_1, \ b_1= d_1$. The constraints on $a_2$, $b_2$, $c_2$, and $d_2$ resulting from the choice of $x_2$ are the same.  Testing has shown that all 16 of the resulting rules will fail for some initial configurations.  \ep

A final note concerns lattices of prime size.
 It has been conjectured, in the case of radius 3, 
 that there may exist rules with the desired behaviour on arbitrary lattices of prime size \cite{Wolz.deOliveira.2008}.  While the case of radius 3 is still open, we can show  
 that the impossibility result  for radius 2 holds even if we restrict the 
 discussion to prime-sized lattices.

\begin{figure}[!h]
 \centering     
 \mbox{  {\scalebox{.45}
{\includegraphics{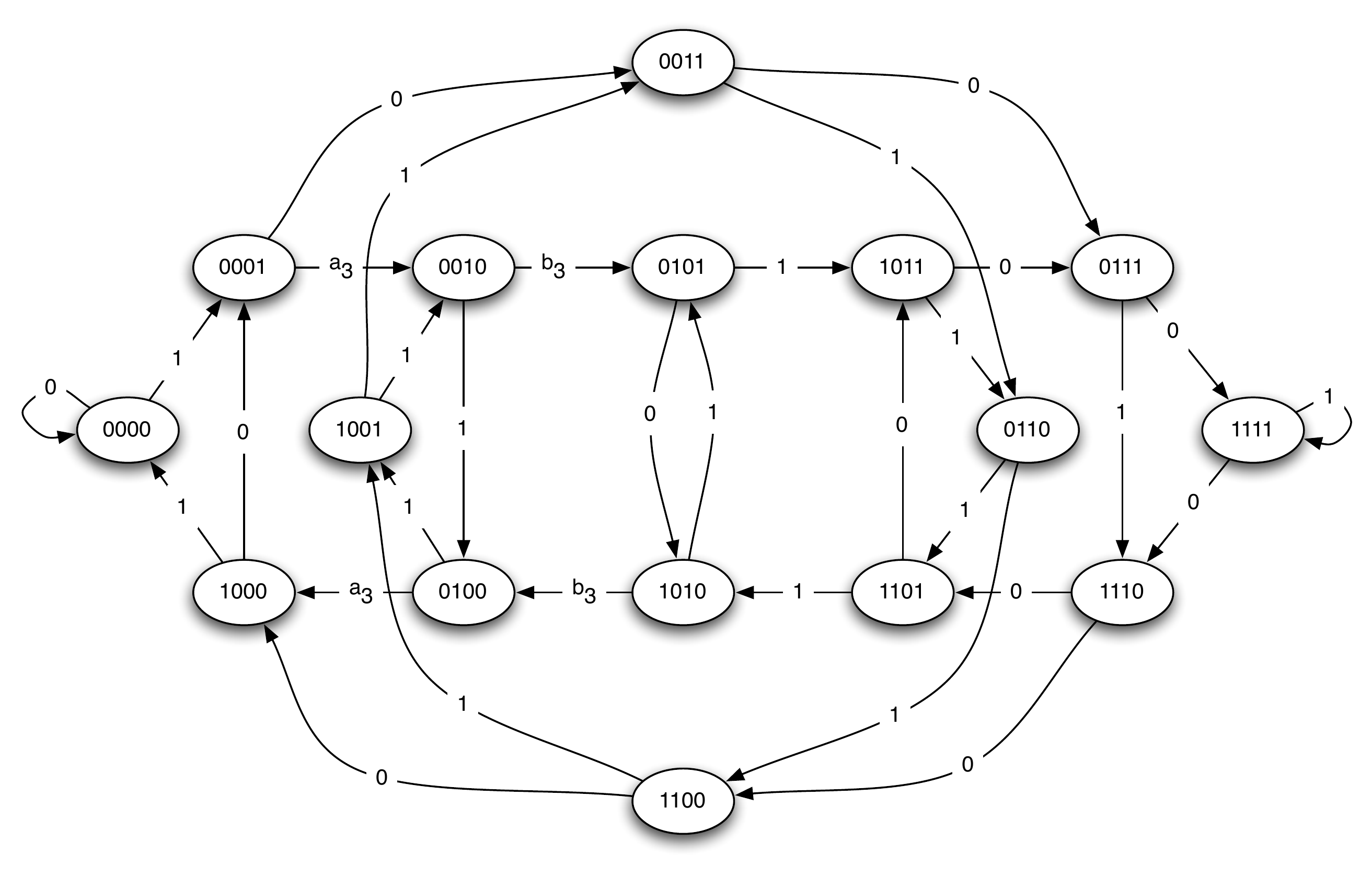}}}
}
\caption{Additional possible perfect rules for radius 2 on prime lattices with $W_2^5$.}\label{r2v3}
\end{figure}

\begin{figure}[!h]
 \centering     
 \mbox{  {\scalebox{.45}
{\includegraphics{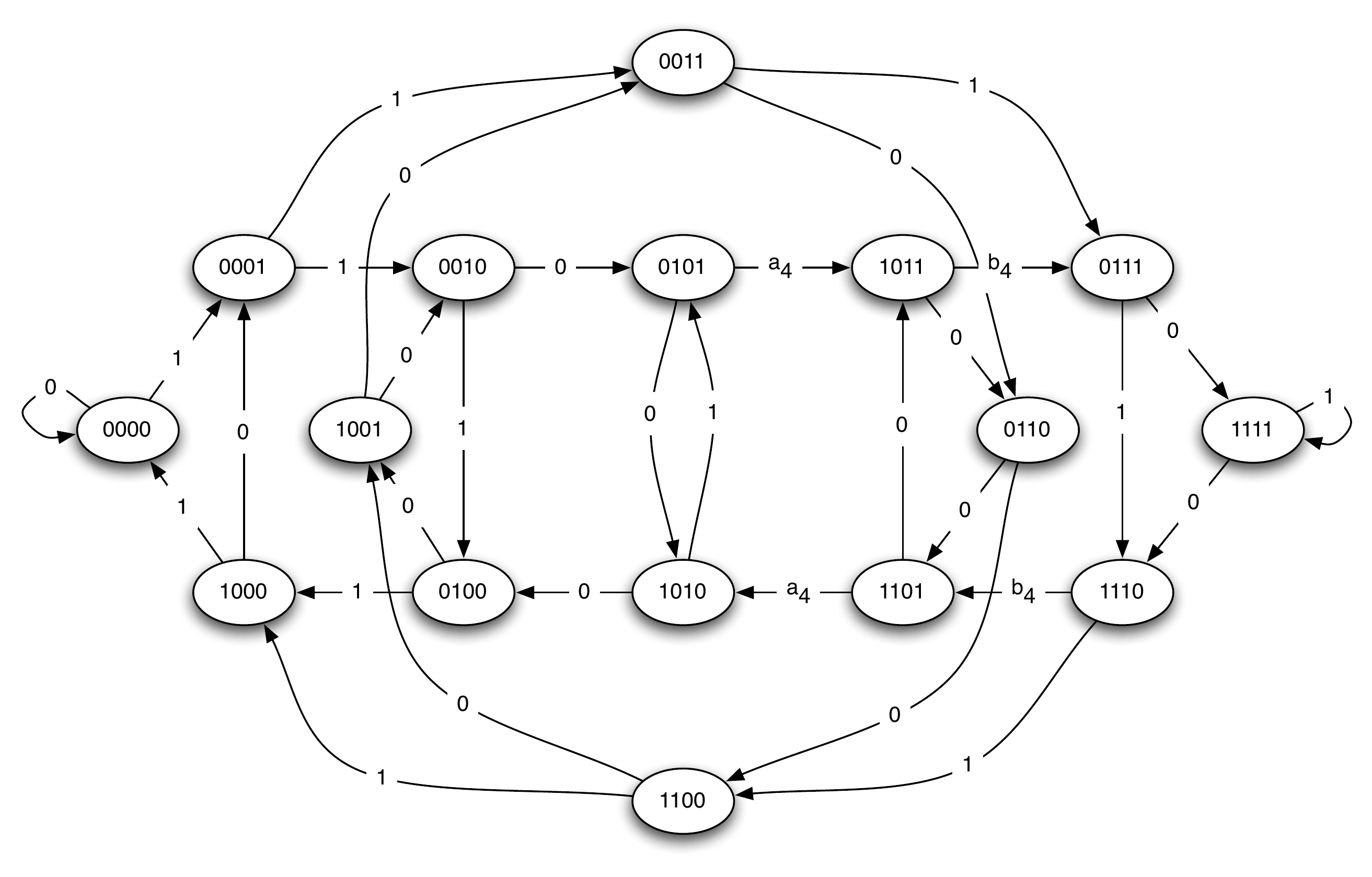}}}
}
\caption{Additional possible perfect rules for radius 2 on prime lattices with $B_2^5$.}\label{r2v4}
\end{figure}

\begin{theorem} There is no   radius 2 rule the always solves the parity problem
even restricting to    lattices of prime size.
\end{theorem}
\pf
 Restricting to prime size lattices, we can no longer use Lemma  \ref{moreconstraints}.  
 This introduces only a few possible extra cases using $B_3^5$ or $W_3^5$. Consider rules containing $B_3^5$. As before, we can eliminate $W_1^5$, $B_2^7$ and $B_3^7$ as pre-images of the 0- and 1- configurations.  In addition, $B_3^5$ conflicts with $W_2^7$ and $W_3^7$ on the edge from $0101$ to $1011$ and with $W_4^7$ on the edge from $0110$ to $1101$.  Hence, we must have $W_1^7$ as the 7-cycle pre-image of the 0-configuration. Now, $B_1^7$ conflicts with $W_1^7$, so we are left with $B_4^7$ as the 7-cycle pre-image of the 1-configuration. Since $B_4^7$ conflicts with $W_3^5$ on the edge from $0100$ to $1001$, we are left with $W_2^5$ as the 5-cycle pre-image of the 0-configuration. These results are illustrated in the graph  of Figure \ref{r2v3}. Proposition \ref{pairs} dictates that the edges labeled $a_3$ must be the same, as will be the edges labeled $b_3$. Now consider  rules containing $W_3^5$.  Similar analysis  shows that must have $B_2^5$, $W_4^7$ and $B_1^7$, as illustrated in Figure \ref{r2v4}. Testing has shown that all 8 of these rules will also fail for some initial configurations of prime length.\ep

\section{A  Perfect Rule with Radius 4}
 
We now describe the construction of a rule with radius 4 having the desired properties: 
parity preservation and convergence.  The intention is to first give the reader an intuitive understanding 
of how the rule works and how it was developed.  Formal proofs will follow in the subsequent section.

\subsection{Rule {\sc BFO}}

The most compact representation of rule {\sc BFO} that we propose for solving the parity problem in any lattice of odd size is given in Figure \ref{comp} and corresponds to rule number:

{\small
\noindent 
12766019579927887748828308783632125137208948629571434199404394002671695991869267727-}

\noindent
{\small 072917454377539194754200976283425175983876539715064584172642413634846720}

\noindent in Wolfram's lexicographic ordering scheme. The figure shows all active transitions (i.e., transitions that change the current state).  However,  it is often easier to explain why and how the rule works using a less compact form, where pairs of rules can be made explicit.  This form of the rule is given in Figure \ref{ruleBFO}; we will be referring to this representation  in the remainder of this section.

\begin{figure}[h]  
\centering
 \mbox{  
\begin{tabular}{|c|c|}
\hline
{\bf Neighbourhood}  & {\bf Output} \\
{\bf configurations}  &{\bf bit} \\
\hline
$*$\hspace{-2pt} 1\hspace{-2pt}   1\hspace{-2pt} 1\hspace{-2pt} {\bf 0}\hspace{-2pt} 0\hspace{-2pt}   $*$\hspace{-2pt} $*$\hspace{-2pt} $*$ & 1 \\
1\hspace{-2pt}   1\hspace{-2pt}   1\hspace{-2pt} 0\hspace{-2pt} {\bf 0}\hspace{-2pt} $*$\hspace{-2pt} $*$\hspace{-2pt} $*$\hspace{-2pt} $*$ & 1 \\
$*$\hspace{-2pt} 0\hspace{-2pt}   0\hspace{-2pt} 1\hspace{-2pt} {\bf 0}\hspace{-2pt} 0\hspace{-2pt}   $*$\hspace{-2pt} $*$\hspace{-2pt} $*$ & 1 \\
0\hspace{-2pt}   0\hspace{-2pt}   1\hspace{-2pt} 0\hspace{-2pt} {\bf 0}\hspace{-2pt} $*$\hspace{-2pt} $*$\hspace{-2pt} $*$\hspace{-2pt} $*$ & 1 \\
$*$\hspace{-2pt} $*$\hspace{-2pt} 0\hspace{-2pt} 1\hspace{-2pt} {\bf 0}\hspace{-2pt} 1\hspace{-2pt}   0\hspace{-2pt}   0\hspace{-2pt}   $*$ & 1 \\
\hline
1\hspace{-2pt} 1\hspace{-2pt} 1\hspace{-2pt} 0\hspace{-2pt} {\bf 1}\hspace{-2pt} $*$\hspace{-2pt} $*$\hspace{-2pt} $*$\hspace{-2pt} $*$ & 0 \\
$*$\hspace{-2pt} 0\hspace{-2pt} 1\hspace{-2pt} 0\hspace{-2pt} {\bf 1}\hspace{-2pt} $*$\hspace{-2pt} 0\hspace{-2pt} $*$\hspace{-2pt} $*$ & 0 \\
$*$\hspace{-2pt} $*$\hspace{-2pt} 0\hspace{-2pt} 1\hspace{-2pt} {\bf 1}\hspace{-2pt} 0\hspace{-2pt} $*$\hspace{-2pt} $*$\hspace{-2pt} $*$ & 0 \\
$*$\hspace{-2pt} $*$\hspace{-2pt} $*$\hspace{-2pt} 1\hspace{-2pt} {\bf 1}\hspace{-2pt} 0\hspace{-2pt} 1\hspace{-2pt} 1\hspace{-2pt} 0 & 0 \\
$*$\hspace{-2pt} $*$\hspace{-2pt} $*$\hspace{-2pt} 0\hspace{-2pt} {\bf 1}\hspace{-2pt} 1\hspace{-2pt} 0\hspace{-2pt} $*$\hspace{-2pt} $*$ & 0 \\
$*$\hspace{-2pt} $*$\hspace{-2pt} $*$\hspace{-2pt} $*$\hspace{-2pt} {\bf 1}\hspace{-2pt} 1\hspace{-2pt} 0\hspace{-2pt} 1\hspace{-2pt} $*$ & 0 \\
\hline
\end{tabular}
 }
\caption{Minimised rule BFO (the asterisk $*$ refers to any value).}\label{comp}
\end{figure}

\begin{figure}[tbh]  
 \mbox{  
\begin{tabular}{|c|c|c|}
\hline
{\bf Name} & {\bf Neighbourhood}  & {\bf Output} \\
 & {\bf configurations}  &{\bf bit} \\
\hline
$T_{1}$:  & $*$\hspace{-2pt} 1\hspace{-2pt} 1\hspace{-2pt} 1\hspace{-2pt} {\bf 0}\hspace{-2pt} 0\hspace{-2pt} $*$\hspace{-2pt} $*$\hspace{-2pt} $*$ & 1 \\
$T_{2}$:  & 1\hspace{-2pt} 1\hspace{-2pt} 1\hspace{-2pt} 0\hspace{-2pt} {\bf 0}\hspace{-2pt} $*$\hspace{-2pt} $*$\hspace{-2pt} $*$\hspace{-2pt} $*$ & 1 \\
$T_{3}$:  & $*$\hspace{-2pt} 0\hspace{-2pt} 0\hspace{-2pt} 1\hspace{-2pt} {\bf 0}\hspace{-2pt} 0\hspace{-2pt} $*$\hspace{-2pt} $*$\hspace{-2pt} $*$ & 1 \\
$T_{4}$:  & 0\hspace{-2pt} 0\hspace{-2pt} 1\hspace{-2pt} 0\hspace{-2pt} {\bf 0}\hspace{-2pt} $*$\hspace{-2pt} $*$\hspace{-2pt} $*$\hspace{-2pt} $*$ & 1 \\
$T_{7}$:  & $*$\hspace{-2pt} $*$\hspace{-2pt} 0\hspace{-2pt} 1\hspace{-2pt} {\bf 0}\hspace{-2pt} 1\hspace{-2pt} 0\hspace{-2pt} 0\hspace{-2pt} $*$ & 1 \\
\hline
$T_{5}$:  & $*$\hspace{-2pt} $*$\hspace{-2pt} $*$\hspace{-2pt} 0\hspace{-2pt} {\bf 1}\hspace{-2pt} 1\hspace{-2pt} 0\hspace{-2pt} $*$\hspace{-2pt} $*$ & 0 \\
$T_{6}$:  & $*$\hspace{-2pt} $*$\hspace{-2pt} 0\hspace{-2pt} 1\hspace{-2pt} {\bf 1}\hspace{-2pt} 0\hspace{-2pt} $*$\hspace{-2pt} $*$\hspace{-2pt} $*$ & 0 \\
$T_{8}$:  & $*$\hspace{-2pt} 0\hspace{-2pt} 1\hspace{-2pt} 0\hspace{-2pt} {\bf 1}\hspace{-2pt} 0\hspace{-2pt} 0\hspace{-2pt} $*$\hspace{-2pt} $*$ & 0 \\
$T_{9}$:  & $*$\hspace{-2pt} $*$\hspace{-2pt} $*$\hspace{-2pt} 1\hspace{-2pt} {\bf 1}\hspace{-2pt} 1\hspace{-2pt} 0\hspace{-2pt} 1\hspace{-2pt} $*$ & 0 \\
$T_{10}$: & 1\hspace{-2pt} 1\hspace{-2pt} 1\hspace{-2pt} 0\hspace{-2pt} {\bf 1}\hspace{-2pt} 0\hspace{-2pt} $*$\hspace{-2pt} $*$\hspace{-2pt} $*$ & 0 \\
$T_{11}$: & 1\hspace{-2pt} 1\hspace{-2pt} 1\hspace{-2pt} 0\hspace{-2pt} {\bf 1}\hspace{-2pt} 1\hspace{-2pt} 1\hspace{-2pt} $*$\hspace{-2pt} $*$ & 0 \\
$T_{12}$: & $*$\hspace{-2pt} $*$\hspace{-2pt} 1\hspace{-2pt} 1\hspace{-2pt} {\bf 1}\hspace{-2pt} 0\hspace{-2pt} 1\hspace{-2pt} 1\hspace{-2pt} 0 & 0 \\
\hline
\end{tabular}
 }
 \mbox{   
\begin{tabular}{|c|r|}
 \hline
{\bf Pair} & {\bf Behaviour} \\
 \hline
$T_{1},T_{2}   $& Rightward growth of 1-blocks \\
$T_{3},T_{4}$&  Rightward growth of 1-blocks     \\
$T_{5},T_{6}$&  Annihilation of 11       \\
$T_{7},T_{8} $&  Local shift    \\
$T_{9},T_{10} $& Leftward growth of 0-blocks  \\
$T_{9},T_{11} $  & Start of  0-growth \\
$T_{9},T_{12}$ & Local adjustment    \\
 \hline
\end{tabular}
 }
\caption{Active rule transitions (left) and behaviour of combinations of rules (right).}\label{ruleBFO}
\end{figure}

We now describe  the intended  behaviour of the rule  before proving its correctness.
Consider an initial configuration $X^0$ as being formed by blocks  $b_i$ of consecutive 1s 
separated by blocks $w_i$ of consecutive 0s:
$X^0=(b^0_1,w^0_1,b^0_2,w^0_2,$
$\cdots, b^0_k,w^0_k \cdots)$.
The idea of our construction is to have a block  $b_i$ 
of 1s propagate to the right, two cells per iteration, until a stopping condition or convergence has been reached.
Such propagation {\em might} result in merging the block with the next  $b_{i+1}$
(if the corresponding  $w_i$ is of even size).
When the merger does not occur (because $|w_i|$ is odd  or due to some other condition),  
there will be a propagation
of 0s to the left, led by a block of the form  (01). Such counter-propagation {\em might} result in the
total annihilation of the block of 1s. Otherwise, it will result in the creation of a single 1 surrounded by 0s, which will start propagating to the right again.  We will show that such behaviour reduces the number of blocks, eventually converging to 
an homogenous configuration.

We now describe     some properties of the rule that can easily  be derived by
construction and that give an intuition for the reasons for the behaviour described above.
 Note that, by construction, the rule's transitions always occur in pairs; in other words,  
 whenever a transition occurs in a cell, another transition occurs in its neighbourhood.
  It is useful to 
 describe the behaviour of each pair and we will also use these pairs to prove that parity is being preserved. 
\medbreak
 
\noindent --- {\bf Rightward growth of 1-blocks.} 
A  singleton 1   grows to the right, by  two  1s at each step,  if it is preceded and followed by at least two 0s, 
as prescribed by  transitions $T_{3}$ and $T_{4}$.
A block of three or more 1s   grows to the right if it is followed (on the right) by at least two 0s. 
This behaviour is created by the pair of transitions $T_{1}$ and $T_{2}$.

\noindent --- {\bf Annihilation of   pairs of 1s.}  As a consequence of transitions $T_{5}$ and $T_{6}$, 
an isolated pair of 1s 
is always   eliminated.

\noindent --- {\bf Leftward  growth of 0-blocks.}  A (01) block moves to the left, 
leading a growing block of 0s (at a growth rate of two 0s per step)  if there are
 at least three 1s to its left  and one of the following:  
 $(i)$   at least three 1s to the right of the 0
  (the growth is obtained by the pair of transitions, ($T_{9},T_{11}$); or
$(ii)$  at least one 0 to its right (due to $T_{9}$ and $T_{10}$). 
Note that the pair ($T_{9},T_{11}$) {\em starts} the growth of a  0-block, while the pair ($T_{9},T_{10}$)
  {\em continues} the growth as far as possible.
 
\noindent  ---  {\bf Local Shift.}  A (101) block is  transformed into (110) if there are
a 0 on its left and at least two 0s on its right
(combination of transitions $T_{7}$ and $T_{8}$).

\noindent  ---   {\bf  Local Adjustment.}   Finally, if a (0110) block is preceded by at least three 1s  that is, 
 $(\ldots 1110110 \ldots )$ occurs, in order to avoid parity errors due to the annihilation of the pair of 1s, we force the creation of  a solid block of 0s to the right of the existing block of 1s with 
 transition pair $(T_{9},T_{12})$, so that  $(\ldots 1110110 \ldots)$  becomes $(\ldots 1000000 \ldots)$.

\begin{figure}[!h]
 \centering     
 \mbox{  {\scalebox{.45}
{\includegraphics{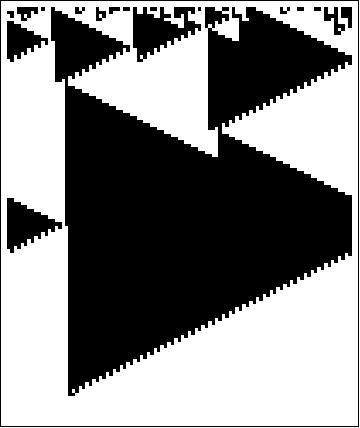}}}
}
 \mbox{  {\scalebox{.45}
{\includegraphics{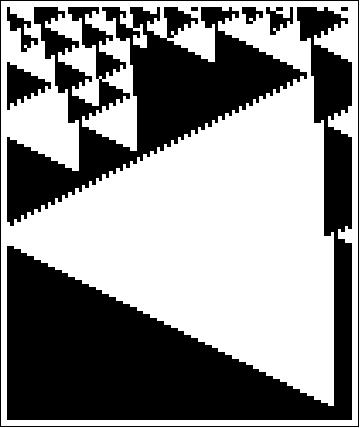}}}
}
\caption{Evolution of rule BFO for even parity (left) and odd parity (right). A black cell  corresponds to 1, a
white cell  corresponds to 0. The initial configuration is at the top and time goes downward.}\label{BFO}
\end{figure}

Examples of the evolution of the rule are given in Figure \ref{BFO}.  

  \begin{figure}
\centering
\scalebox{.5}{\includegraphics{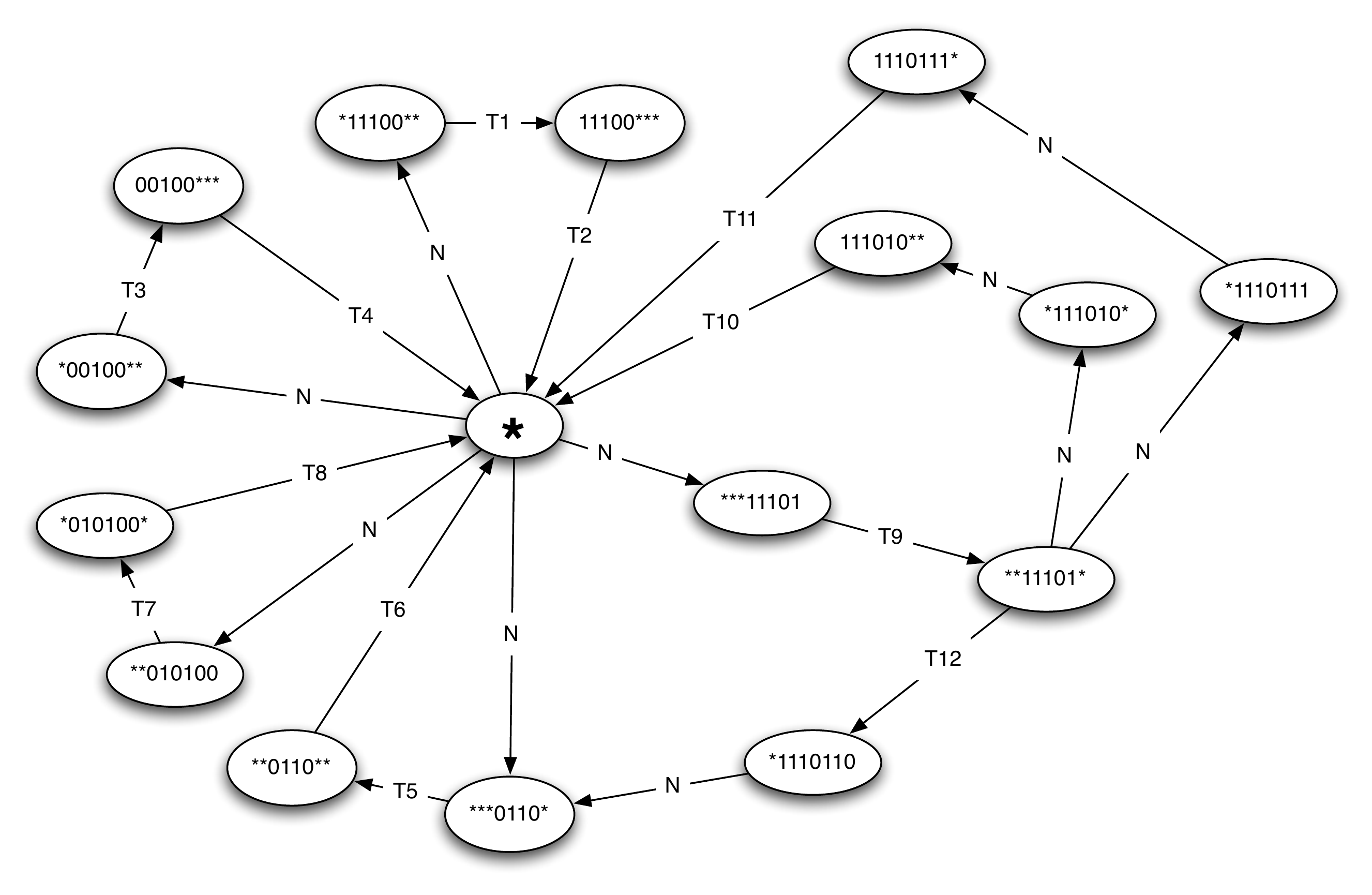}}	
\caption{\footnotesize Reduced transitional de Bruijn graph for rule {\sc BFO}.}\label{dBBFO}
\end{figure}

\subsection{Correctness}\label{sec:proofs}
  
In order to show that the rule we have constructed (or in fact any rule) performs a perfect parity check, we must prove that  it preserves parity at every iteration, and that it   always converges in finite time to an homogeneous configuration.  
We begin with the proof of parity conservation.

\subsubsection{Parity Preservation}

A rule   preserves the parity of a configuration if active transitions always come in pairs.  That is, given a local rule $f$ of radius $r$, and any configuration $(x_0,\dots,x_{n-1})$, the number of times that $f(x_{i-r},\dots,x_i,\dots,x_{i+r})\neq x_i$ is even. 
 To show that our rule does indeed have this property, we will use a modification of the de Bruijn graph.  

Given a configuration $X=(x_1,\dots x_n)$, one can determine the next iteration, $F(X)$ by reading the edge labels as one traverses the graph from $(x_{i-4}$ $ \dots$ $x_i\dots x_{i+3})$ to $(x_{i-3}$ $\dots$  $x_i\dots x_{i+4})$.  Since we are considering    circular configurations, the
 traversal of the de Bruijn graph will result in a closed loop.
For our purposes, we are not interested in what the actual output is, only if an active transition has occurred.  In keeping with that, we are, in fact, only interested in the parts of the graph where such transitions occur. 
Since a  de Bruijn graph for a function of radius 4 can be quite unwieldy,   we define a {\em reduced transitional de Bruijn graph} 
modifying the   standard   de Bruijn  graph as follows.
  First, we   label  the edges with $T_i$ (indicating one of our 12 active state transitions), or $N$, meaning  no transition.  Second, we draw only those parts of the graph connected to transitions, reducing the rest of the graph to a single node, denoted by an asterisk.  Also,  where there is no conflict, several nodes leading to or from the same transition are depicted as one,  using the notation of the previous section; for example, there is an edge from node $(*11100**)$ to 
  $(11100***)$, because of the various state transitions entailed by $T_{1}$, given all possible values for the $*$ symbols. Furthermore, we   ensure that  all nodes in the reduced transitional de Bruijn graph are distinct for all values of the $*$ symbol. 
Finally, for any nodes occurring explicitly in the graph, all adjacent edges are represented, whether they correspond to a transition or not.
It is easy to see that parity preservation can be detected from the reduced transitional 
de Bruijn graph for a given rule; more precisely:

\begin{lemma} A rule is parity preserving if and only if  any cycle in its reduced 
transitional de Bruijn graph contains an even number of edges labeled with 
some transition $T_i$.
\end{lemma}
\pf
Since the output of a circular CA is given by the edge labels of a cycle in its de Bruijn graph, we need only count the transitions to verify parity preservation.  Furthermore, the reduced graph compresses only parts of the graphs where no transitions occur.  Hence, if the are no cycles
containing an odd number of transitions in the reduced graph, there can be no configurations leading to an odd number of transitions and vice versa.
\ep

 By inspecting the 
transitional de Bruijn graph for rule {\sc BFO}, and by noticing that     there are no cycles containing an odd number of transitions, we then have:

\begin{theorem}
Rule {\em BFO} is parity preserving.
\label{parity}\end{theorem}

\subsubsection{Convergence}

We now turn our attention to the more challenging problem of proving that this rule will converge under any conditions.  We can think of any CA configuration as an alternating sequence of blocks of 0s and blocks of 1s of varying lengths.  We will show that {\sc BFO} eventually converges by showing that its only fixed points are the homogeneous configurations and, furthermore, that any change in the configuration will lead, in a finite number of iterations, to a reduction in the overall number of blocks.
Our first lemma shows that  every non-homogeneous configuration is changing.  
\begin{lemma} The only fixed points of rule {\em BFO} are the homogeneous configurations. \label{grow}\end{lemma}
\pf A fixed-point configuration cannot contain pairs of 1s, since rule pair $(T_5,T_6)$ would apply.  It cannot contain a block of three or more 1s since  $(T_1,T_2)$ would apply  if it is followed by two or more 0s, and a pair containing $T_9$ would apply if it is followed by only one 0. Therefore, a non-homogeneous fixed-point configuration could only contain isolated 1s but the odd length of the configuration would imply that we must have at least two consecutive 0s, hence the sub-configuration 0100 must occur.  If this is preceded by a 0, $(T_3,T_4)$ apply. If it is preceded by a 01, $(T_7,T_8)$ apply.
\ep

  We now wish to show that every transition pair will eventually lead to a reduction in the total number of blocks. We begin with the transition pairs for which this is immediate.
    
    \begin{lemma} 
    Transition pairs ($T_{5},T_{6}$), ($T_{7},T_{8}$) and quadruplet ($T_{9},T_{12},T_{5},T_{6}$)
     lead to block reduction in a single step.\end{lemma}

\begin{lemma}
Transition pair ($T_{9},T_{10}$) leads to block reduction in a finite number of steps.
 \end{lemma}
\pf
Rule pair $T_{9}:f(***1{\bf 1} 101*)=0$ and $T_{10}:f(1110  {\bf 1} 0***)=0 $ causes the leftward growth of 0-blocks.
While a single application of this pair    maintains the number of blocks,
 it leads (possibly through a repeated application of the pair) to an eventual   block reduction through either 
 an annihilation  or the creation of a single 1:
\begin{eqnarray*}
 01111010\cdots &\leadsto& 01101000 \cdots  \textrm{ by rules $T_{9}$ and $T_{10}$}\\ 
                     & \leadsto& 00001000 \cdots  \textrm{ by rules $T_{5}$ and  $T_{6}$}\\
 0111010 \cdots &\leadsto& 0101000 \cdots \textrm{ by rules $T_{9}$ and $T_{10}$}\\ 
 &\leadsto& 0110000 \cdots \textrm{ by rules $T_{7}$ and  $T_{8}$}\\
 &\leadsto& 0000000 \cdots \textrm{ by rules $T_{5}$ and  $T_{6}$}\\
\end{eqnarray*}

\noindent Notice that ($T_{9},T_{10}$) leads to reducing the number of blocks by either two or four,  depending on the parity of the block of 1s on which it is acting. Also note that, even though it is possible for the leading block of 1s to have shrunk from the left side, while the $(T_{9}$,$T_{10})$ pair is reducing it from the right, one of these two situations will still be reached, since, on the left, the 1s can only be eliminated one at a time.
 \ep

 \begin{lemma} 
 Transition pairs ($T_{1},T_{2}$), and ($T_{3},T_{4}$) lead to either  reduction or maintenance of the number of blocks.
 \end{lemma} 
 \pf 
 Both pairs
 $T_{1}:f(*111 {\bf  0} 0*** )=1$, $T_{2}:f(1110 {\bf 0} ****)=1$,
 and $T_{3}:f(*001 {\bf 0} 0***)=1$,  $T_{4}:f(0010 {\bf 0} ****)=1$  are responsible for the
 rightward growth of 1-blocks. In fact, they
 grow a block of 1s until it either merges with the next block or an isolated 0 preceded by three or more 1s is created. 
 At this point, one of the following transition sets will apply:  ($T_{9},T_{10}$) if the isolated 0   is followed by 10, ($T_{9},T_{12},T_{5},T_{6}$) if it is followed by 110, and ($T_{9},T_{11}$) if it is followed by three or more
1s.  We have already seen that the first two cases lead to block reduction, only the latter case 
can maintain block numbers.
\ep

\begin{lemma}Transition pair $(T_9,T_{11})$ leads to  reduction or maintenance of the number of blocks 
in a finite number of steps.\end{lemma}
\pf  
The pair  $T_{9}:f(***1{\bf 1} 101*)=0$,
$T_{11}:f(***1  {\bf 1} 1011)=0 $ is responsible for the start of  0-growth.
This  is the only rule pair that initially increases the number of blocks.  It is the beginning of the growth of 0s from an isolated 0 surround by three or more 1s on either side. Once this growth has begun, it is continued by the $(T_9,T_{10})$ pair until the number of blocks has been returned to its original size or is reduced by two.
\ep
 
Note that in the proof above, if the original number of blocks is maintained, it is because the transitions have produced an isolated 1 with two or more 0s on either side which then begins to grow to the right either merging with the block on the right (and thus reducing the total number of blocks) or creating an isolated 0 which then begins to grow left. What we wish to avoid is  a CA which evolves to a periodic configuration of 1s growing until only an isolated 0 remains and then shrinking back to an isolated 1 which then regrows.  We call this pattern of growing and shrinking {\em  the accordion effect}.  In the next lemma, we show that the accordion effect cannot occur on lattices of odd length.

\begin{figure}[tbh]
 \centering     
 \mbox{  {\scalebox{.45}
{\includegraphics{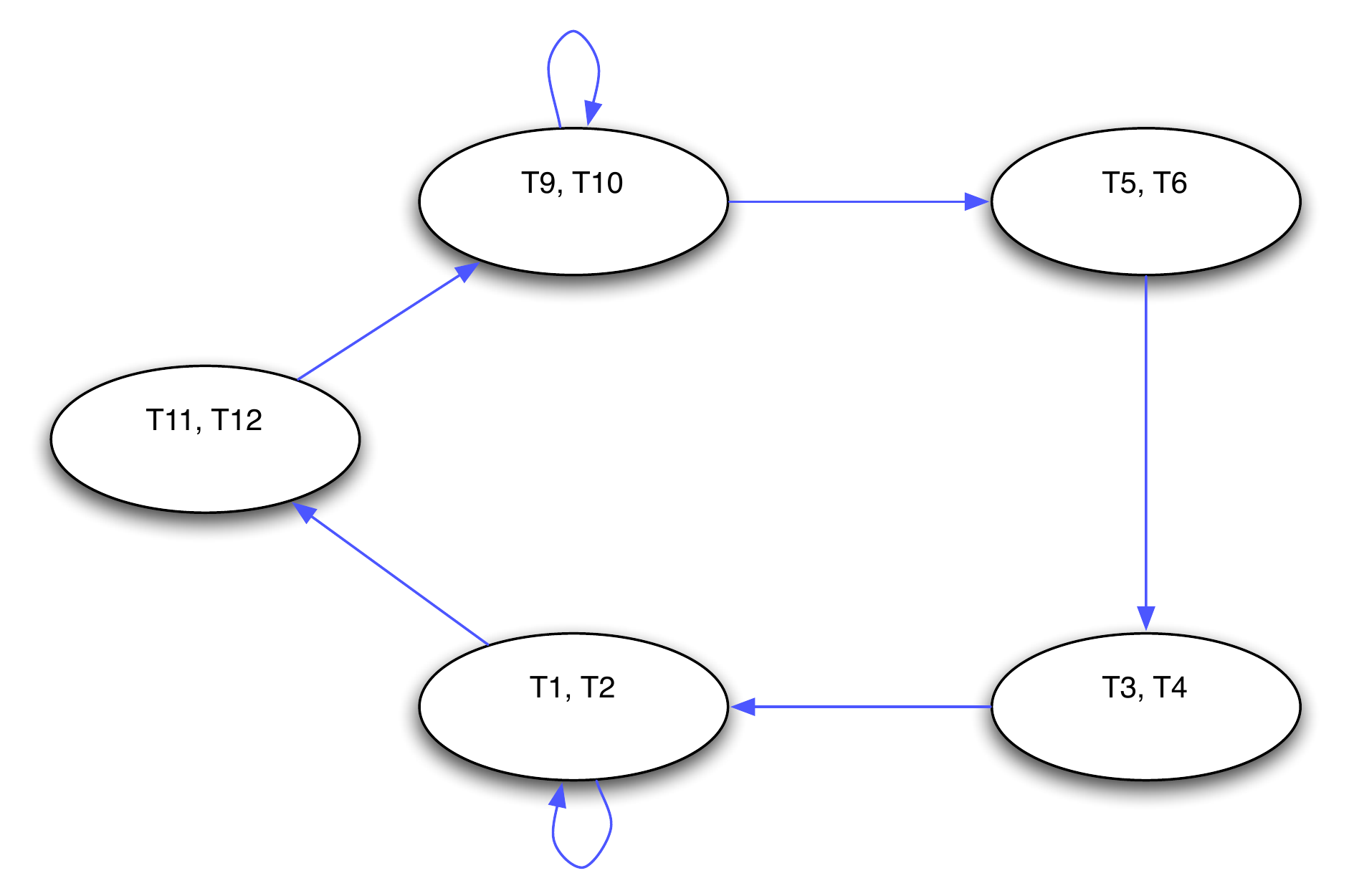}}}
}
\caption{Accordion loop.}\label{accordion}
\end{figure}

\begin{lemma} The accordion effect can only occur on lattices of even length.\end{lemma}
\pf
Figure \ref{accordion} shows the transition pairs involved in a cycle of growing and shrinking.    If we think of this loop as starting from the $(T_9, T_{11})$ node, a block of 1s is shrinking until the ($T_5$, $T_6$) node is reached and then beginning with the $(T_3, T_4)$ pair, a block of 1s begins to grow.  Let us assume that the entire lattice is perpetually in some stage of this cycle. We make several observations: 
\begin{itemize}
\item The 1-blocks start off having odd length.
\item The 0-blocks start off having even length.
\item In order for a block of 1s to regrow, it must have even length at the end of the shrinking process.
\item  In order for a 0-block to  grow, it must have odd length at the end of the growth of 1s. 
\end{itemize}

\noindent  Since the entire configuration is in this process, it is made up of sub-configurations of two forms: $b_i w_i$, a block of 1s followed by a block of 0s when the 1s are growing; or $b_i01w_i$ a block of 1s separated by a 01 from a block of 0s when the block of 1s is shrinking.  Consider a block of 0s, $w_1$ followed by a block of 1s, $b_2$.
If $w_1$ has odd length, then it must not change lengths again before the growth of 0s restarts otherwise when the block of  1s on its left, $b_1$  grows, it will merge with $b_2$.  This means that $(T_{9},T_{11})$ must be applied on the left before $(T_{5},T_{6})$ can be applied on the right.  Hence $b_2$ will shrink by 1 before the 1s have finished shrinking on the right.  Since we will need to have an even number of 1s at that time, we must now have an odd number of 1s.  In other words, a block of 0s of odd length is always followed by a block of 1s of odd length.
Now assume that $b_1$ has even length.  In this case it must change sizes before the regrowth of 1s is complete, so  $(T_{5},T_{6})$ must be applies on the right before $(T_{9},T_{11})$ is applied on the left again.  Hence $w_2$ must already have even length.
Taken together, we see that for the accordion effect to endure in perpetuity, the CA must have even length.\ep
 
    \begin{figure}[tbh]
 \centering     
 \mbox{  {\scalebox{.45}
{\includegraphics{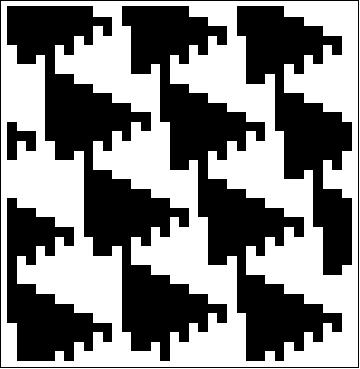}}}
}
 \mbox{  {\scalebox{.45}
{\includegraphics{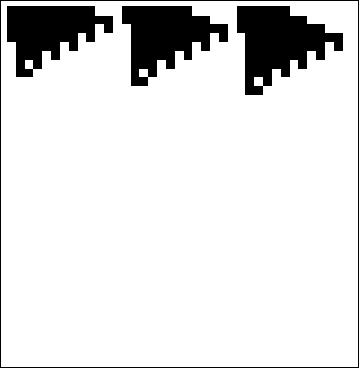}}}
}
\caption{The accordion effect on an even-sized lattice (left)  and the successful resolution of an odd-sized lattice (right).}\label{failing-even}
\end{figure}

\noindent Finally, taking these various lemmas together, we have the heart of our convergence proof.

 \begin{lemma} From any non-homogeneous configuration, the total number of blocks   decreases in finite time.\label{shrink}\end{lemma}
 \pf From the previous lemmas, we see that only rule pair $(T_{9}, T_{11})$ increases the number of blocks and that within a finite number of steps, this increase is resolved. Our contention is that the accordion effect is the only way to maintain block numbers.  Consider a $(T_{9}, T_{11})$ pair occurring anywhere in our CA. If the leading block of 1s is odd at the end of the execution of the $(T_{9}, T_{10})$ pairs, then the number of blocks had decreased by two.  If it is even, the 1s will begin to grow right.  If the  0-block created by the execution of $(T_{9}, T_{11})$ is unchanged from the right, then it has even length and the regrowth of 1s  will result in a merger of 1s and a reduction in the number of blocks.    Now the only way for the 0-block to have changed on the right is if the 1-block on its right had shrunk due to the application of $(T_{9}, T_{10})$ or  $(T_{9}, T_{11})$.  If the reduction in the 1-block was not initiated by the $(T_{9}, T_{11})$ pair, then real reduction in the number of blocks has occurred.  If it was initiated by  $(T_{9}, T_{11})$, then block reduction can only be prevented if another  $(T_{9}, T_{11})$ pair is being executed to its right.  Arguing in this way, we see that reduction in total block number can only be avoided if the CA is experiencing the accordion effect which can only occur in even-sized lattices.
\ep

\noindent  From Theorem \ref{parity} and   Lemmas \ref{grow} and   \ref{shrink}, we obtain:
 \begin{theorem} 
 Given a CA of odd length, rule {\em BFO}   converges to all 1s if the initial configuration has odd parity and to all 0s if it has even parity.\end{theorem}

 \section{Concluding Remarks}
 
 In this paper, we have established upper and lower bounds on the radius of  rules 
that solve  the parity problem by showing that there exists a rule of radius 4 which converges to all 1s if the initial configuration
 is odd, and to all 0s if it is even and, further, by proving that this problem is unsolvable by rules of radius 2, even with the less strict condition of prime-sized lattices. 
 The corresponding questions for radius 3 remain open.  However, we have developed tools in this paper that should be helpful in solving this latter problem as well.
 
It is clear by now, how painstaking the task of designing a CA rule can be, let alone the formal proof of its correctness. To some extent, the process reminds us of similar programming efforts on simple, pre-modern computational models, such as Turing machines. And in this sense, we are still indeed at this point in history, when programming cellular automata.

Since our main motivation for addressing the parity problem is not conscribed to it, one may ask how generalisable our experience herein could be to related problems, including the parity problem for radius 3, as well as other related computational problems for CAs. 
It is tempting to think of the possibility of implementing a high-level programming approach that would automatically generate the state transitions of a CA rule, given the kinds of notions we have used, such as the growth of blocks of a given size in a given direction, the annihilation of blocks of given kind, etc. Even if this form of programming, so to speak, {\em by patterns}, does not solve the problem of designing a rule (the target algorithm), at least it would help its high-level conception, and its implementation in terms of the required state transitions.

As a methodological note, it is worth mentioning that it was  demanding in practice to resort to computational aids to complement the formal efforts. After all, the details involved in rule design are so many that it is quite easy to overlook some of them. This  turned out to be essential in the present case for fine tuning our design in its origin. Such an interplay between formal and computational methods also came into play for devising the most compact representation of the BFO rule, as shown in the paper, for enumerating all cycles of length 7 in the de Bruijn graph of radius 2, and for the evaluation of the radius 2 rules that had retained potential for being perfect solvers of the parity problem, by not violating the constraints derived in the proof, at their various stages of development.

\section*{Acknowledgements}

This work has been partially supported by the Natural Sciences and Engineering Research Council of Canada (NSERC), by Prof. Flocchini's University Research Chair, and by a grant provided to P.P.B.O., by MackPesquisa - Fundo Mackenzie de Pesquisa.

\mbox{}
 \bibliographystyle{eptcs}
 \bibliography{automata}

 \end{document}